# ET PROBES, NODES, AND LANDBASES: a Proposed Galactic Communications Architecture and Implied Search Strategies

**JOHN GERTZ** Foundation for Investing in Research on SETI Science and Technology (FIRSST). Correspondence: Zorro Productions, 125 University Avenue, Suite 101, Berkeley, CA, 94710. Tel: (510) 548-8700 USA

**email** jgertz@firsst.org

Land-based beacons, information laden probes sent into our solar system, and more distal communication nodes have each been proposed as the most likely means by which we might be contacted by ET. Each method, considered in isolation from ET's point of view, has limitations and flaws. An overarching galactic communication architecture that tethers together probes, nodes, and land bases is proposed as a better overall solution. From this more efficient construct flows several conclusions: (a) Earth has been thoroughly surveilled, (b) Earth will be contacted in due course, (c) seti beyond half the distance that Earth's EM has reached (~35-50 LY) is futile, and (d) the very quiescence of the galaxy paradoxically implies that that Drake's N = many, and that there is a system of galactic governance. Search strategies are proposed to detect the described probe-node-land based communications network.

## 1 INTRODUCTION

ET might be motivated to communicate with Earth in order to trade information; or ET might wish to disseminate its values, beliefs, history, art, religion, and so forth, without regard to receiving anything in return; or ET's main purpose might be to enlist us into a galactic alliance for the wellbeing of all member civilizations, i.e., law and order may be ET's prime imperative.

Any or all of these goals might be achieved by direct civilization-to-civilization communication by means of, for example, laser or radio transmissions. Such civilization-to-civilization transmissions are sometimes referred to in the literature as "beacons," and the search for them (including the search for signals not intended for us, but acquired by us through eavesdropping) constitutes the classic and still most common seti research protocol [1, 2]. However, there are many reasons why ET might not choose to communicate with beacons. For example, a civilization must simultaneously deploy many beacons, each directed at a likely target star for eons in order to happen upon a receiver that might eventually turn to it. Alternatively, if ET's transmitter sequentially targets thousands of stars, while Earth's receiver sequentially targets thousands of stars then the chances of them aligning in time are nearly impossible unless they are targeting the exact same set of stars (why would they?), and even then the chances of alignment are tiny during any one duty cycle. Additionally, there is the immense problem of receiving a return message. ET must either have an all-sky-all-the-time system or have a receiver dedicated to each and every star on its target list, since it will have no *a priori* idea which star might harbor a listening civilization or in which duty cycle its signal might be received, or how long the receiving civilization might take to decode the message and decide upon its response. However, this is only the start of the sending civilization's problems. The two civilizations must be able to communicate in some fundamental way. If one speaks in words, as we understand the term, but the other speaks in colors or bee-like waggle dances, the chance of communicating anything more than "we also exist" is small. If ET is vastly more advanced than us (a statistical likelihood, since what are the chances that, in the billions of years in which it might have evolved, they also are in their first century of electronic technology?), then communication may be pointless from ET's point of view. After all, would we feel the need to discuss literature or mathematics with *Australopithecus aferensis*?

There is then the problem of the immense danger facing any civilization that decided to signal. Beacons are deadly dangerous (do not try this at home). A beacon would give away ET's exact location, as would Earth with its reply. Just in the signal strength and type, ET will give away much about its level of advancement. An embedded message may have very unintended consequences. For example, the Voyager plaque showing a naked man and woman with the man's hand raised in a sign of peace, might convey that our species has no thorns, horns, shells or other obvious means of defense, and is therefore weak. Who knows but that a raised hand might have a similar meaning to ET as what a raised middle finger means in American culture. As good looking as we think we are, ET might find our visage repulsive, vaguely resembling some vermin on its planet. Our plaque sent in peace might actually invite a pre-emptive attack.

If the beacon is very bright and very complicated, the receiving civilization may fear to respond. If ET communicates with a flux that is a million times more powerful than our best transmitter, Arecibo, dare we respond with Arecibo and thereby let ET know just how feeble we are? If we try to bluff ET into believing that we are a mighty species with awesome powers, but that we merely choose not to use our own gargantuan ra-dio transmitter, do we perhaps invite the pre-emptive annihilation we are trying to avert because they would fear us [3]? The dangers may be so profound that ET civilizations that employ beacons (i.e., engage in METI) may not achieve the longevity necessary for likely detection by Earth because they are soon exterminated by others.

These and many other problems can be solved if ET simply sends a message to us in a probe [4,5,6,7,8,9], which is defined here as ET-derivative objects sent to any solar system for the purposes of surveillance and/or communication with its resident technological species. Probes offer a variety of advantages over beacons. As AI machines they can hide their planet of origin, or reveal it as judgment might warrant. For all but nearby stars, the per bit cost of the information payload (e.g., contained on something like a small thumb drive) is vastly cheaper than the cost to transmit the same information via continuously broadcasting beacons [10,11]. It is hard to be more specific than this, since an exact comparison would need to consider such unknown factors as the mass of the probe, its speed, the gravity of its launch site and minimum escape velocity from its solar system versus the width of the informationally comparable beacon's beam at the point of reception (determining the amount of wasted flux) and the length of time a beacon must continuously broadcast before being received.

Probes would be able to surveil at close range, gathering data on their target solar systems even in the absence of communication with a technological species. Probes nullify the limiting effect of Drake's *L*, the length of time that an intelligent civilization transmits beacons, since probes might long survive their progenitor civilizations. However, these and other advantages of probes are for naught if ET wishes to receive a message in return. Probes might pile up uselessly, waiting for such time, as ever, as a technological society might evolve in the solar system of their presence. They could spend those eons doing surveillance and research, but without a gargantuan transmitter they would not be able to relay their findings home.

This author [12,13] has argued that the problems inherent in probes can be addressed by a system of nodes, defined as probes that are not necessarily located within our solar system but that are explicitly designed to communicate among themselves, as well as with technological civilizations within their area of service. The fact that beacons are so inherently dangerous that no truly intelligent species would dare to deploy them may actually have given rise to a system of communication beacons that are offset from the civilizations associated with them. Early ET civilizations may have sent beacons to orbit nearby stars, reasoning that if the respondent launches an attack it will be against a decoy. Likewise, the first civilizations that responded to those beacons would do so from their similarly offset transmitters. As a result, a proto-nodal communication system may be of the same age as the very first ET-to-ET transmissions.

However, as this author has previously described them, nodes also have distinct drawbacks. They would have only a limited capacity to surveil Earth if the nearest one is located at an interstellar distance. At even close interstellar distances, Earth's EM leakage damps down to incoherence from which not much useful information might be gleaned other than perhaps that they are of artificial derivation [14].

Here it will be argued that the best solution may be a network of ET probes, nodes, and land-based systems tethered together into a comprehensive communications architecture. In terms of hierarchical computing power, one might, by analogy only, imagine a probe as having the power of a laptop; a node as having the power of a supercomputer; while a land base might host a server farm and multiple quantum supercomputers. For conceptual purposes only, probes might orbit every star, and therefore be spaced at ~3-6 light years (LY) in our region of the galaxy; nodes might be positioned at 25 LY intervals, and land-based member civilizations might be located at an average distance of 250 LY from one another. It also might be that nodes orbit most or all stars for the purpose of repeating and amplifying signals. Perhaps most of these are sub-nodes in that, while being signal repeaters, they otherwise lack comprehensive analytical abilities.

## 2 PROBES

### 2.1 Probes as Surveillance Systems

Flyby probes. In the first instance, nodes or land bases might send simple flyby probes through every stellar system for the purposes of surveilling them. Breakthrough Starshot (BTS) is designing just such a system to surveil nearby stars within the current century [15]. One of the greatest challenges facing BTS is in designing a robust communications system for sending information back to Earth. However, assuming that ET's probe is travelling at a low enough speed, it might dispense with this complexity by simply recording data as it flies through a planetary system on a trajectory that approaches its host star closely and accurately enough to be bent into a new trajectory aimed directly at any node of convenience, that not necessarily being the node that sent it. The flyby probe would simply upload its data as it approaches that next node. That node would in turn send that data on to adjacent nodes at the speed of light.

Speed is of the essence—i.e., slow speed. A major objection to the above is that by not transmitting the information it has gathered at the time of the flyby the probe has greatly lengthened the time of its mission, sacrificing light speed for physical travel time. This, then, would be a good moment to digress to a consideration of the question of time. In order to understand the proposed galactic communications architecture, it is important to understand that probes and nodes would likely be AI machines, for which human time scales are almost meaningless. Sol is not among the first generation of stars that might have borne technological life. There is no known reason why the first ET civilization in the galaxy might not have arisen more than five billion years ago (bya) [16]. ET's entire library might contain data on hundreds of billions of stars, and have catalogued untold millions or billions of life-bearing planets. ET might have made contact with thousands or millions of technological societies and maintained the parochial libraries of those civilizations indefinitely, irrespective of whether any particular civilization persists or has perished.

In deep time, the fact that a probe might take some additional tens of thousands of years to complete its mission may be of trivial concern. Initial flybys might have been sent past a molten Earth 4 bya and found it to be lifeless, but the same probe might have found life on Mars. As a consequence, perhaps the node sent follow up probes at 100 million year intervals, soon discovering that Mars had turned lifeless, while Earth blossomed with microbial life. It would not have been until about 2.2 – 2.5 bya that it might have detected atmospheric molecular oxygen and then presumed that multicellular life might have arisen. On such a 100 million year time scale, the speed of any one flyby mission would not be a major consid-eration. In fact, slow speed would be much preferred, because (a) it would cost much less in energy to accelerate a probe to the desired speed; (b) the slower the speed, the less severe the damage from collisions with interstellar dust or micrometeorites (though cosmic rays would remain a problem—but one that is soluble through shielding and/or self-correcting data redundancy); (c) less energy would be required to decelerate upon arrival at the target system using solar radiation and solar wind (it could use the same solar wind and radiation to reaccelerate out of our solar system); (d) at slower speeds the probe's trajectory can be more easily bent as it approaches the star; and (e) more data might be gathered at slower flyby speeds.

Orbiting probes. Once Earth had been determined to have entered its multicellular life form phase, one or more permanently orbiting probes might have been sent. If ET has a good idea of the average time between multicellularity and the emergence of a technological species, it might time its probe launches appropriately. Again, slow speed is of the essence to allow for the probe to be able to efficiently enter into an orbit without the need to bring immense amounts of fuel along for deceleration. Permanent probes would be outfitted with interstellar class communication receivers and transmitters. Given that the signal, be it radio or optical, would likely be monochromatic and of an exactly known frequency to the receiving node, it can be of relatively modest power.

Just because we are being surveilled does not mean that we live in a zoo. The probes hypothesized here are compatible with, but different from, those hypothesized in the well-known "zoo hypothesis," in that probes might be here to study us, but they also are hypothesized to have the capacity to communicate with us at some point. We are not necessarily merely a nature preserve. Upon a probe's detection of our EM leakage, it might require a significant amount of time for it to analyze those signals. Yet much more time would be required if the probe merely reformats and amplifies our leakage, focusing and transmitting it onwards to a node or land base for complete analysis and to communicate instructions. This process might take centuries or millennia. However, the net result is that when we do hear from our local probe(s) or node, it might be in fluent English, or some other terrestrial language.

### 2.2 Probes as "Lurkers"

David Brin has postulated that probes reside in our solar system now, but actively decline to make their presence known [17]. He calls them "lurkers," in effect, spies. Brin offers a number of possible reasons for their refusal to make contact, among which are (a) that they have seen our nightly news and either fear or loathe us; (b) they are waiting for us to pass some developmental milestone by which we might prove our worthiness; (c) they follow a strict policy of non-interference (i.e., the zoo hypothesis); or (d) they have no need to trade information with us as they can simply steal it from our abundant EM communications streams, with no intention of reciprocating.

However, there are other possibilities:

- Probes are studying us in order to understand how best to make contact. Probes would only have had about a century to observe us through our EM leakage. We have studied dolphins and whales for decades and still cannot speak with them. It may simply take a lot of analysis to understand us.

- The local probe(s) may not have the onboard capacity to accomplish a full analysis of our civilization, nor the authority to make contact. It may be passing information back to a node or land based command and control center for a final determination. In such an event, contact will be made in due course, but that contact may be a century or more into our future.

- The local probe(s) or node is signaling us now, but we simply have not yet made the detection, or, we may actually have made a detection, but thrown it out as a hard to interpret transient or as RFI (see below).

- A probe might appropriate all the knowledge it could want from our EM leakage. However, the current hypothesis may mean that Earth has a crucial role to play in the galactic communications architecture. In exchange for full membership in the galactic club, Earth might be required to dedicate itself to building and launching nodes and probes. In effect, Earth might assume responsibility for the maintenance of the local communication network. This, of course, would also be in Earth's own interest.

### 2.3 Where are they?

If our solar system contains many probes, in the spirit of Fermi, it should be asked why we cannot readily see evidence of them. The very simple answer is that probes are probably fairly small, possibly no larger than an automobile, and perhaps as small as a tennis ball or smaller. In our vast solar system, they would remain unnoticed were their discovery solely reliant upon serendipity. We might only find them if they are actively signaling to Earth, or if we catch them in the act of signaling to a node [18].

## 3 NODES

### 3.1 What's the rush?

As with probes, high speed is counterproductive. If nodes are launched into free orbit around the center of the Milky Way (MW), they would seem to be most functional when travelling at about the same orbital speed as the star systems they serve. If nodes orbit stars (including possibly our own), then, as with probes, a slower entry speed into our solar system is necessary in order facilitate deceleration, and thereafter maneuver into an orbit. Moreover, a node that moves too quickly will be flung out of the MW.

### 3.2 The Features of Nodes

This author has elsewhere envisioned nodes as having commonalities with cell phone towers, the Internet, an escrow service that ensures fair information exchange among civilizations, and a lending library [12].

### 3.3 Nodes as a Dating App

To the above, this author would now add that nodes may provide the physical backbone for software that may be most akin to a dating app. There has been the tacit assumption among seti researchers that ET will transmit to us a uniform body of knowledge. However, nodes may no more represent a homogenous galactic civilization than does the Internet. There may, in fact, be no Encyclopedia Galactica (EG) embodying a single galactic culture. Nodes might instead resemble a dating site that marries together civilizations with enough in common to intelligibly communicate. Our local node might recommend specific civilizations to us. EG might actually be a collection of parochial encyclopedias. If nodes help determine, for example, that a given civilization might have enough in common with Earth, it would send them our Wikipedia or Encyclopedia Britannica, and send us their equivalent. It might also translate from one language to another, or give us the Rosetta stone tools to enable us to do this ourselves. Then if mutually desired, it could put each of us in direct communication with the other through the galactic node system, masking the exact location of each. This feature of the galactic communication network might allow for the following features:

- We would be free to adopt the religion or art of one civilization, while rejecting that of another civilization.

- Instead of relying on the node's judgment, we might be free to examine ET profiles, choosing for ourselves which civilizations we wish to further study. Our node might offer only a thumbnail at first, and then would download on request as much as we wanted from that civilization, subject perhaps to the rule that the node nearest that civilization would be authorized to download our full set of information as well (albeit, much later as limited by the speed of light). There might be no point in putting us in contact with a cuttlefish-like color communicating civilization, or one that is a billion years in advance of us. By digging into our node's databank, we would, in effect, have a virtual conversation with any ET civilization of interest. The actual civilization could be quite remote, say, 50,000 LY away. It might also be the case that some, or even all, of the civilizations of most interest to us are no longer extant, or that we have made virtual contact with a civilization much earlier in its history, and that the same civilization is currently too advanced for meaningful direct communication.

- Crucially, the proposed probe, node, and land base galactic architecture offers all involved anonymity, and therefore security. Even if civilizations A and B decide to communicate in real time, limited only by light speed, they will never know each other's locations. This is because their respective messages might go through dozens or hundreds of nodes, of whom only the first and last know the positions of the respective civilizations. Anonymity may be a vital and intended feature of the system not just for the safety and security of individual civilizations, but also for the whole of the galaxy. Anonymity would serve as a break on would be attempts to form aggressive alliances or empires.

- The proposed architecture allows for multiple networks, just as many countries have competing cell phone systems or one might choose among various search engines.

## 4 LAND-BASED COMMAND AND CONTROL CENTERS

It may be that probes report to nodes, and nodes report to land-based command and control centers. Whereas probes and nodes are in all likelihood AI machines, the beings that inhabit the land bases may be life forms of recognizable organic chemistry, or they may be robotic AI beings, having been made by but then replaced their natural predecessors. Land bases within the proposed architecture would not be the capitals of empires any more than a server farm hosting an Internet cloud is an emperor. There may be no ruler anywhere in the galaxy, even though, as argued above, there must be rules. If simple enough, the rules can run on autopilot with rewards (membership in the galactic architecture) and punishments (complete annihilation) that are simple to understand and administer.

Von Neumann replicators are interesting theoretical constructs [19], but they may not exist. In addition to possibly being galactically illegal, they may be too difficult to construct in practice. Imagine a replicator on an asteroid. It strains the imagination as to how it might fabricate computer chips in the absence of any manufacturing facility; or how it would mine all necessary materials, say, the lithium and cobalt that might be necessary for batteries. It might be far easier for land bases to be the mining and manufacturing hub for the probes and nodes that would be needed to service its local area.

Land bases might also be the repository of the deepest store of knowledge, and have the greatest computing sophistication. Land bases should be considered broadly as having the natural resources to build probes, nodes, and supercomputers and endowed with the intelligence to organize this, be that intelligence carbon based, AI or other. Small bodies, such as asteroids, or artificial bodies, such as space stations, are not precluded. If the proposed architecture is reliant upon land bases, then our participation in the galactic order is necessary. This will be Earth's greatest bargaining chip. ET will not need us for our knowledge, and in fact may get annoyed or bored if we try to impress it with how many prime numbers we know. As for E=Mc2, they have heard that one too many times before also. They may be mildly interested in Beethoven or Jackson Pollack, but their real need is for us to assist in maintaining the galactic communications network in Earth's immediate neighborhood. Consequently, ET probes will not "lurk" indefinitely. The galactic architecture needs us, and we are not therefore helplessly dependent upon ET's merciful altruism.

## 5 THE GALACTIC COMMUNICATION ARCHECTECTURE AS LAWGIVER AND POLICE FORCE

Tipler has argued that we are alone, at least in the MW if not in the entire visible universe, because if we were not alone, an earlier aggressive civilization would surely have created a von Neumann replicator, and that one solitary machine would proceed to exponentially recreate itself *ad nauseum* until it had gobbled up all matter in the galaxy [20]. Alternatively, had even one ordinary carbon-based life form set out to colonize the galaxy, due to its exponential growth, it too would have acted like a horde of locusts, per the classic interpretation of Fermi's Paradox. Either way, the result is a tragedy of the common. Since, obviously, this has not happened, it is deduced that we must be alone. However, there are at least two other possible solutions to the conundrum. It may be that Drake's N does not = 1 (us), but that N = 2; us plus one other, a "berserker" civilization, that, like the first queen bee that emerges from her cell and immediately kills all other queens that would otherwise have hatched after her, this berserker civilization (or its AI creations) will destroy us as soon as it detects our EM leakage [21].

There is, though, a third possibility, namely that the apparent quietude of the galaxy paradoxically leads to the deduction that N = at least a few and perhaps many, because a lack of colonization must mean that there is a system of galactic governance in place, regulated by probes, nodes, and land bases, that vigilantly guards against the type of tragedy of the common implied by Tipler or the berserker hypothesis. In practice, seti research is based on the premise that N = many, and therefore it presupposes a system of governance.

Haley, inventing the term "metalaw," proposed a galactic rule book governing relations among its technological civilizations [22]. Others either elaborated upon or criticized the pu-tative corpus of laws [23,24,25]. Proponents argue in favor of a universal principle of "do unto others," and "equality," while opponents tend to criticize on grounds that metalaw is anthropomorphic with scant application to beings far more advanced or far different from humans, much less robotic life. In this author's view, laws must exist, but they must be equally applicable to AI as well as naturally evolved life. This set of rules might include: (a) severe limitations to colonization, perhaps limited to a civilization's immediate non-biological neighborhood, or even to one nearby star, and only then when necessitated by the imminent death of one's own star; (b) outlawing of von Neumann replicators; and (c) outlawing of aggression against other civilizations, probes, nodes, and land bases. One might dispense with all the bodies of law that might govern non-aggressive behavior, brotherhood, equality and the like, since under the proposed architecture, no civilization knows the whereabouts of any other, and therefore can do it no harm. Since colonization is forbidden, there would be no border disputes. The galaxy's entire rule book may consist of two words variously translatable as: "stay still" or "keep away."

The penalty for noncompliance would presumably be planetary destruction. It is speculated that the probes/nodes/land bases would have the means of disposing of aggressive civilizations as automatically as our immune system deals with pathogens, without the need for judge or jury, as we would understand those terms. This should not surprise us, as we ourselves are only a few years away from deploying autonomous automobiles and battlefield weapons whose algorithms will be fully empowered to make life and death ethical decisions. That said, probes could act as an early warning system that would presumably be able to determine far in advance whether a civilization is likely to go rogue, and might be able to take deterring actions short of annihilation.

Alternatively, and bleakly, probes might be currently surveilling us with the intent of deciding whether to engage with us—or to destroy us. They may now be analyzing the data (e.g., violent newscasts versus feel-good movies like *Forrest Gump*). The local probe(s) may be streaming data back to a local node, which in turn is passing it back to a command and control land base, all in a grand determination of whether the galaxy should welcome us, leave us alone for the moment while more data is gathered, perhaps over the next several millennia, or destroy Earth forthwith. An algorithm that is as incapable of caring or mercy as we are when we pull a weed from our garden may even now be determining Earth's fate. It may be that a local probe, placed within the Solar System eons ago, is laced with a self-replicating toxin or nano-grey-goo that can be instructed from afar to destroy life on Earth. If Judgment Day is well neigh, it will be by act of computer code rather than of a god.

## 6 SEARCH STRATEGIES

The proposed probe/node/land base galactic communications architecture hypothesis leads to a set of recommended search strategies. Most of these assume that one or more probes or nodes are transmitting to Earth now, but there might also be viable strategies by which communications between an intra-solar system probe or node and a more remote node might be intercepted.

### 6.1 Emphasize Nearby Stars

Targeted star searches are recommended for nearby stars that have had time to receive our EM leakage and to have responded, i.e., stars ~<35 – 50 LY. Our local node may be in orbit around any of them. Reasonable solutions to the Drake Equation when applied to biologically based and transmitting civilizations yield a distance to the nearest such civilization as being much further than this. However, it is very reasonable to assume that nodes may be much closer than the nearest organic land based civilization, and that nearby stars are therefore prime candidates so long as they are closer than ~<35 – 50 LY. Beyond that, it is recommended that the predominant seti search paradigm be reversed, whereby stars are explicitly targeted and foreground probes or nodes would only be detected by serendipity. The new paradigm would search for local probes and nodes, but be fully able to detect, by serendipity, background and distant beacons, say by preferentially targeting local objects such as asteroids or the area around the sun, when they are transiting the plane of the MW.

### 6.2 Trade High Sensitivity for Wider Field of View (FOV)

Targeted star searches generally use a narrow beam with a small FOV and high sensitivity, as the signal they would detect is presumed to be weak because it comes from a great interstellar distance. The preferred strategy for detecting intra-solar system probes/nodes would be to employ a wide FOV, since an intra-solar system probe might be found in any direction. In exchange for this wide FOV, high sensitivity can be sacrificed, since a probe in our own solar system is presumed to be transmitting at a power that any reasonably capable telescope might detect. Since telescopes generally have a fixed FOV, the effective recommendation is to use the existing FOV to mosaic areas of interest, sacrificing integration time, and with it sensitivity, to do so.

### 6.3 Enlist Amateurs

Amateur radio and optical astronomers might be recruited. If the sought-after signal is local, and therefore robust, amateurs might be able to successfully detect it. Given the plethora of previous seti and non-seti sky surveys, it is unlikely that an amateur would detect a persistent signal that had somehow been missed. However, if a probe or node broadcasts to Earth very intermittently, say once a decade, perhaps from the surface of an asteroid, and then only for a short period of time, then an all-hands-on-deck strategy may be necessary in order to detect its signal.

### 6.4 Where to Look within Our Own Solar System

A probe that is transmitting to Earth could be orbiting the sun in any plane, forcing us to survey 360 degrees of sky. However, we can make some reasonable surmises that would narrow the search:

- Preferentially target the asteroid belt between Mars and Jupiter [2]. A probe might have landed on an asteroid where it might dig in for protection against micrometeorites and radiation. There would be abundant sunlight for PV generation and useful materials for self-repair and the build out of capabilities.

- Take a special look at Earth's only known Trojan asteroid, the ~30 km. wide 2010TK, located at Lagrange point, L4.

- Use the Large Synoptic Survey Telescope (LSST) to detect probes and nodes. When the LSST becomes fully operational in 2022 it will conduct the most thorough visible wavelength survey of the sky ever. It will have an exceptionally wide FOV, and is designed to repeatedly survey the entire sky visible to it. It will be ideally suited for optical seti (oseti) as traditionally conceived as detecting a laser beacon originating from a remote star system. All that would be needed is a further set of algorithms capable of detecting a laser signal in non-sidereal motion as it moves through our own solar system, either in free orbit around the Sun or Earth, or in solar orbit astride an object such as an asteroid or moon.

- Preferentially target Lagrange points, especially the more stable L4 and L5, including the Lagrange points around both Earth and the other planets, on the theory that probes would require less energy to maintain orbital stability.

- The best solution for a probe entering our solar system may be for it to enter into a highly eccentric orbit around our Sun (a node might prefer a tight circular orbit around the Sun, considered below). This would obviate the need for the additional gear necessary for it to land on an asteroid or moon, and it would require the least amount of energy relative to what would be needed to either land on a solar system object, such as a moon or asteroid, or enter into a planet mimicking near-circular orbit either around the Earth or the Sun. A simple probe, one with no landing gear, and with a small communications system capable of reaching only the nearest stars, might not require the materials for build out and self-repair afforded by asteroids. There would also be the added benefit to the probe of being able to store the Sun's energy as it closely approaches the Sun. It might also assume an orbit that allows it to closely fly by Earth (the one obviously biological planet) once or twice in each orbit for closer surveillance. This suggests that instead of searching the 360 x 360 = 129,600 square degrees of sky, we might restrict the search to a more manageable ~100 - 200 square degrees of sky around the Sun, a region never yet explored. Radio searches can be accomplished any time during the day, while laser searches might be conducted just before sunrise and after sunset, as well as during solar eclipses. Regarding the latter, archival photos of total eclipses should be re-examined for lasers. It is possible that a laser was caught in the act but ignored as a lens or pixel malfunction. Were the very first attempt to look for radio signals near the Sun to succeed, it would either imply that probes in highly elliptical orbits must be legion, or that there is only one or a few probes, but they are in tight, almost circular, orbits around the Sun. A probe in a highly eccentric orbit would only travel a small fraction of its orbit within the 100 – 200 square degrees closest to the Sun. If probes are not legion, and the first observation fails, follow-up observations should be planned at regular intervals, certainly not less than once per year.

- If orbiting the Sun in a highly eccentric orbit, a probe might be found in any plane, necessitating a 360-degree search. However, the search area might be reduced somewhat by deemphasizing or ignoring altogether the ecliptic, reasoning that of all possible planes of orbit, the probe would be least likely to be found in the debris-strewn ecliptic plane, through which it might otherwise be making hundreds of thousands or millions of transits in the course of deep time. This strategy flies in the face of a strategy predicated on the hypothesis that local probes would choose to sit on asteroids. Astronomers might choose the paradigm they think more promising.

- Search for life within our own outer Solar System which closely resembles Earth based life. Obviously, this is already a major goal of NASA and other space agencies. If life on Enceladus or Europa proves to be Earth-like this may be evidence for panspermia. Panspermia between the Earth and the outer Solar System, in turn, might be taken as *prima facie* evidence for the past or present existence of probes that seeded them in common.

## 6.5 Probes and Nodes as Transients

Seti researchers should accept the possibility that ET signals may not be sidereally stationary. The greatest bane of seti research is the transient, a signal that mimics what might be expected from an extra-terrestrial beacon, but, if found in real time, disappears before confirmation, or if found offline, is absent when researches return to the same fixed location. The 1977 WOW signal is the most famous example, but there are many more. When a transient is recorded that is not obviously RFI, instead of assuming that it might have come from a star in the FOV, researchers should also examine the possibility that the source may be a foreground probe within our own solar system. They should determine whether an asteroid or moon was also in the FOV at the time of detection, and then observe at the then current position of that object. If no such object was in the FOV at the time of original observation, or if that confirmatory observation is negative, they should test to determine whether the source might have been a free orbiting probe. If the original detection was made in real time or not too long thereafter, the telescope should slowly slew in a spiral around the point of the detected transient.

## 6.6 Eavesdropping on Probe-to-Node or Node-to-Node Communications

If a probe is not communicating with Earth, but in communication with nodes in nearby star systems (i.e., it is currently in lurker mode), there may still be ways in which we might eavesdrop on its signals:

- Whereas probes might be most efficiently placed in an eccentric solar orbit, nodes might be best placed in close, nearly circular solar orbits. Of course, there would be abundant energy available to it there. However, there may be more important reasons for nodes to locate themselves in close solar orbits. This would allow a sister node orbiting a nearby star to easily locate the node and thereby allow for the use of a very narrow, probably laser and probably phase arrayed beam. For example, a node in a solar orbit at 0.1 AU would unfailingly receive 0.2 wide beamed communications from its sister if that beam is centered on the Sun. This saves transmitter energy, while having the added benefit of being undetectable by life bearing planets, except in the rare instances that the planet happens to transits that 0.2 AU space as seen from the perspective of the transmitting node. For that reason, both the Sun and nearby stars should be viewed at times when the Earth transits between them. However, the better way to detect such a node, unless it is intentionally transmitting to Earth, would be by placing one or more receivers in solar orbits, in the hope of catching a node in the act of transmitting. Because the probe or node may be actively avoiding detection, the preferred orbit of our spacecraft would be far away from the Earth, and would sample both the Sun and the direction exactly opposite the Sun.

- It is recommended to observe at the meridian at and around midnight. A probe or node in a tight orbit around the Sun might be caught in the act of receiving a transmission from a node on the other side of Earth. This technique is particularly recommended when a nearby star is in that position.

- We might observe in a direction that is roughly in exact opposition to nearby stars. If a lurking probe is located on one side of Earth and is signaling to a node on the other, we might succeed in eavesdropping.

- Future space missions like LUCY, an upcoming NASA mission to several of Jupiter's Trojan asteroids, as well as currently flying missions to asteroids, Hayabusa2 (Japanese) and Oris-Rex (NASA) should train their radio receivers on their targeted asteroids to intercept possible probe-node communications. The asteroids should be observed from all directions, not assuming that its probe's signal is directed toward Earth.

- Similarly, archival data from the Saturn probe, Cassini, as well as asteroid probes, Hayabusa1, Rosetta and Dawn, and Jupiter probe Galileo should be re-examined for evidence of unaccounted for transmissions from asteroids or moons. In the future, any NASA or other space agency probe to solar system objects should be outfitted with both radio and optical receivers designed to detect ET signals.

## 7 CONCLUSIONS

A comprehensive search of our own solar system for evidence of ET probes may have been stymied by a wholly unwarranted association with UFOs [26]. Probes and nodes may be distinguished from UFOs along a number of parameters. UFOs have been the subject of investigation for decades, but not one shred of credible evidence has emerged in support of their existence. To date, virtually no explicit search for probes or nodes has been undertaken. UFOs are often assumed to be piloted by living beings, the so-called "little green men." Probes are hypothesized to be artificial constructions impervious to the needs of carbon life forms, such as ourselves. Therefore, they would not require the type of massive life support systems that carbon beings would require for multi-generational interstellar flight. UFOs are presumed to have entered the atmosphere. Probes are presumed to surveil Earth from space, from where they can readily monitor our EM transmissions. No special expertise is required to study UFOs, only a video camera and a butterfly net to catch the little green men if they land. Being objects in outer space, the study of probes and nodes lies squarely within the purview of astronomy.

Tens of thousands of individual stars have been surveyed for radio or laser signals of technological origin. Breakthrough Listen is in the process of surveying one million more [27]. Perhaps the time has come to acknowledge the possibility that ET may no more be using star-to-star communications (except where nodes or land bases are orbiting nearby stars) than to be using smoke signals. Instead, perhaps the time has come to devote significant resources toward the comprehensive search of our own solar system for ET communications.